# Highly asymmetric optical properties of $\beta$-Ga$_2$O$_3$ as probed by linear and nonlinear optical excitation spectroscopy

*Jeong Bin Cho, Gunwoo Jung, Kyuheon Kim, Jihun Kim, Jung-Hoon Song, Joon Ik Jang[*],*

J. B. Cho, J. Kim, Prof. J. I. Jang

Department of Physics, Sogang University, Seoul, 04107, South Korea

E-mail: jjcoupling@sogang.ac.kr (J. I. Jang)

G. Jung, K. Kim, Prof. J. H. Song

Department of Physics, Kongju National University, Kongju, 32588, South Korea

**Abstract**

β-Ga$_2$O$_3$ is a highly promising semiconductor for a deep ultraviolet emitter owing to its wide bandgap, which significantly varies in the range of 4.49 eV to 4.74 eV due to its optical trirefringence in the monoclinic crystal structure ($C_{2h}$). However, dominant photoluminescence (PL) emissions in β-Ga$_2$O$_3$ occur far below the bandgap, where the underlying PL mechanism has been under intense debate. In order to address this issue, we monitored the most intense PL peaks at 3.0 eV and 3.5 eV at room temperature by employing both linear and nonlinear absorption spectroscopy. The PL of β-Ga$_2$O$_3$ was found to be highly polarized along the (102) direction regardless of input polarization, and to be essentially the same for one-photon absorption (1PA), two-photon absorption (2PA), and three-photon absorption (3PA). However, absorption polarization dependence is quite distinct for 1PA, 2PA, and 3PA, as theoretically explained by optical selection rules for $C_{2h}$ symmetry. Based on the input-output method, 2PA and 3PA coefficients were determined to be $\beta$ = 3.45 cm GW$^{-1}$ and $\gamma$ = 0.013 cm$^3$ GW$^{-2}$ at the excitation wavelength of 460 nm and 550 nm, respectively. These values are several times higher than theoretical values predicted by two-band models, indicating that β-Ga$_2$O$_3$ possesses strong optical nonlinearity with high polarization contrast. The absorption power dependences for 1PA, 2PA, and 3PA indicate that optical excitation is essentially excitonic, but bound to Ga vacancies (3.0 eV) and O vacancies (3.5 eV), which is different from the so-called self-trapped exciton based on the polaron model. Our main thesis is also clearly supported by the dependence of the PL shape on the excitation depth as probed by 2PA depth scan. This implies that densities for the Ga vacancy and the O vacancy vary differently as a function of depth from the excitation surface. A broad set of highly asymmetric optical properties clarified in our work is critical for the understanding of this wide-gap semiconductor and its potential use for ultraviolet sources especially when band-edge emission is realized.

**Main text**

Among other polymorphs of Ga$_2$O$_3$, β-Ga$_2$O$_3$ is most stable and crystallizes in the monoclinic crystal structure ($C_{2h}$), having a wide but highly asymmetric bandgap in the range from 4.49 eV to 4.74 eV.[1,2] This wide-gap semiconducting material has attracted much attention for applications in the field of catalysis, gas sensor, and power electronic devices, as well as for optoelectronic devices in the deep ultraviolet (UV) region, since it has the much wider bandgap than GaN (3.4 eV) and ZnO (3.37 eV).[2–7] In addition, the availability of Ga$_2$O$_3$-Al$_2$O$_3$ alloys [(Al$_x$Ga$_{1-x}$)$_2$O$_3$] can make this material a potential candidate for even deeper ultraviolet (wavelength: λ < 220 nm) solid-state light source that can be used to effectively break or disinfect airborne viruses, such as swine influenza virus H1N1 and severe acute respiratory syndrome coronavirus (SARS-CoV).[8,9] Along with its wide bandgap, the availability of high-quality single crystalline substrates and controllable n-type doping have made β-Ga$_2$O$_3$ a promising material for next-generation power electronic devices. From the perspective of optoelectronic and photonic applications, key optical properties of this material must be understood, which include absorption and emission characteristics in the realm of linear and nonlinear light-matter interactions as a function of excitation wavelength and polarization. Intriguingly, upon optical excitation along the widest bandgap direction of β-Ga$_2$O$_3$, the photoluminescence (PL) emission is typically dictated not by band-edge emission (~ 4.74 eV), but by below-gap transitions in UVA (~3.5 eV) and blue (~3.0 eV) regions.[10,11] Based on density functional calculations, depth-resolved cathodoluminescence (DRCL) spectroscopy, ellipsometry and other methods, several mechanisms have been suggested to explain the UVA/blue emissions of β-Ga$_2$O$_3$, including extrinsic defects such as Ga and/or O vacancies,[7,12–15] and intrinsic self-trapped holes (polarons),[10,16–18] but the issue still remains controversial. Obviously, understanding of these transitions and their relationship with emission mechanisms is crucial for Ga$_2$O$_3$-based photonic devices toward the even deeper UV range. Also, precise understanding of its nonlinear optical properties is important for assessing β-Ga$_2$O$_3$ for novel UV nonlinear optical applications. Although β-Ga$_2$O$_3$ was known to be luminescent under multiphoton excitation,[19] there is no report on the precise quantification of its nonlinear optical properties.

In order to study the basic properties of the below-gap PL, we carried out a series of optical measurements on a ($\bar{2}$01)-oriented β-Ga$_2$O$_3$ single crystal by employing the photoluminescence excitation (PLE) spectroscopy over a broad wavelength range (λ = 220 nm–550 nm). This range covers from linear one-photon absorption (1PA) to nonlinear two-photon absorption (2PA) and even to three-photon absorption (3PA). Regardless of excitation order (1PA, 2PA, or 3PA) and input polarization, our sample exhibited typical UVA/blue PL emissions, being highly polarized along the (102) direction. The origin for the highly polarized PL is likely related to the highly anisotropic valence band.[14] We also probed the absorption polarization dependences of 1PA, 2PA, and 3PA of β-Ga$_2$O$_3$, which show very different optical selection rules in accordance with the $C_{2h}$ crystal symmetry.[20–23] Furthermore, we measured the absolute values of 2PA and 3PA coefficients of β-Ga$_2$O$_3$ when the input polarization was along the (102) direction. The measured values can be explained by the two-band models,[24] but with about 5 times larger scaling constants owing to its unique band structure. This observation demonstrates that β-Ga$_2$O$_3$ has very large optical nonlinearity in spite of its wide bandgap. Nonlinear depth scan was also performed to study the below-gap PL as a function of excitation depth. The drastically varying PL feature clearly shows that the nature of PL is not intrinsic, but extrinsic involving Ga vacancies (V$_{Ga}$) and O vacancies (V$_O$) for the blue and UVA emissions, respectively, which is consistent with the recent DRCL results.[12] We further concluded that these transitions arise most likely from radiative recombination of excitons bound to V$_{Ga}$ and V$_O$, based upon both linear and nonlinear absorption power dependences. In fact, the feasibility of a room-temperature-stable exciton with a large binding energy up to 230 meV was theoretically proposed in this semiconductor.[1] Our comprehensive results imply that β-Ga$_2$O$_3$ is not only a promising UV nonlinear optical material with a drastic polarization control but also a deep-UV emitter, especially when the high-quality synthesis is assured for minimizing the extrinsic defects in order to trigger the band-edge emission.

Figure 1a shows the monoclinic crystal structure of β-Ga$_2$O$_3$, together with our excitation geometry specified, where the propagation direction of the laser beam is along the ($\bar{2}$01) direction of β-Ga$_2$O$_3$. In order to compare with 1PA, the beam focus for 2PA and 3PA was properly chosen to maximize surface excitation. Figure 1b, c, and d display the normalized below-gap PL spectra obtained from our β-Ga$_2$O$_3$ under 1PA, 2PA, and 3PA, respectively, where the polarization vector $\hat{\epsilon} = (l, m, n)$ is parallel to the (102) direction. In this direction, the bandgap is in the range of 4.5 eV–4.6 eV, and therefore, our choice of excitation wavelengths is within the range for 1PA (λ = 240 nm), 2PA (λ = 290 nm), and 3PA (λ = 550 nm), respectively. In fact, we confirmed the corresponding excitation order ($p$) based upon the absorption power dependence: PL counts $\propto I^p$, where $I$ is the input intensity of the incident laser and $p$ = 1.0, 1.5–2.2, and 3.3 for 1PA, 2PA, and 3PA, respectively (Figure S1). In view of the extrinsic defect model, the peak near 350 nm (3.5eV) arises from optical transition associated with V$_O$ levels and the peak near 405 nm (3.0 eV) with V$_{Ga}$ levels.[16,25,26] Based on the overlaid PL spectra (Figure S2), we concluded that the PL shape does not depend on the excitation order, except for more pronounced V$_O$ transition with increasing excitation order (1PA→2PA→3PA). As explained later, this effect arises from a partial contribution from volume excitation, which is typical of nonlinear optical absorption, whereas 1PA is essentially surface excitation. Also, a minor wavy feature in the range of 360 nm–380 nm in the 3PA-induced PL turned out to be an artifact when using the short-pass filters employed for 3PA to block the excitation beam. Our results therefore show that the PL is dictated by the below-gap transitions regardless of 1PA, 2PA, and 3PA. We also found that the spectral feature of the PL does not change when $\hat{\epsilon} = (l, m, n)$ was set to parallel to the (010) direction, which is orthogonal to (102). All of these results imply that there are no other hidden levels specifically active to nonlinear optical 2PA or 3PA and that the characteristic of the PL does not depend on the input polarization but on the symmetry of the $V_O$ and $V_{Ga}$ transitions towards the valence band.

In order to probe the symmetry of the below-gap PL, we examined its polarization dependence when the sample was excited by 1PA, 2PA, and 3PA, respectively. Figure 2a displays the schematic for the measurement of the PL polarization with a linear polarizer placed between the sample and the detector. Here, $\hat{\epsilon} = (l, m, n)$ was fixed along the (102) direction. Intriguingly, we found that both $V_O$ transition and $V_{Ga}$ transition are highly polarized along the (102) direction, independent of the excitation order. For example, the polarization dependences of the $V_{Ga}$ transition (3.0 eV) are plotted in Figure 2b, c, and d for 1PA, 2PA, and 3PA, respectively, where the polar angle is the angle between the (102) direction (crystallographic $a+2c^*$ axis in Figure 1a) and the transmission axis of the polarizer. The corresponding degrees of polarization for 1PA, 2PA, and 3PA calculated using eq. 1 are 0.88, 0.94, and 0.9, respectively;

$$P = (I_{max} - I_{min})/(I_{max} + I_{min}) \qquad (1)$$

where $I_{max}$ and $I_{min}$ are the maximum and minimum PL counts. The case for $\hat{\epsilon} = (l, m, n)$ being parallel to the (010) direction (crystallographic $b$ axis in Figure 1a) is shown in Figure S3, again showing that the PL is polarized along the (102) direction regardless of the excitation order. Therefore, Figure 2 and Figure S3 demonstrate that the PL is always polarized along the (102) direction independent of input polarization and excitation order as well. Although defect-induced PL is typically unpolarized, it can attain certain polarization if the levels associated with the transition are anisotropic in k space.[27–29] In our case, the origin for the polarization arises presumably from the valence-band wave function as both $V_O$ and $V_{Ga}$ transitions commonly share the valence band as the final state.

We now present the results on the optical selection rules[20,21,23] for 1PA, 2PA, and 3PA based on the PL counts recorded as a function of polarization angle $\theta$, which is the angle between the $a+2c^*$ axis and $\hat{\epsilon} = (l, m, n)$. Note that these selection rules correspond to the absorption polarization dependence for each excitation order, whereas the results in Figure 2 correspond to the emission polarization dependence. Figure 3a illustrates the schematic for the measurement of the absorption polarization dependence, where $\theta$ was varied by simply rotating the sample. We confirmed that the selection rules for both $V_O$ and $V_{Ga}$ transitions are exactly same as shown in Figure S4 (1PA), Figure S5 (2PA), and Figure S6 (3PA), respectively. The measured absorption polarization dependence for 1PA is plotted in Figure 3b (black dots), superimposed with the theoretical selection rules for 1PA in the $C_{2h}$ structure (red) given by

$$a\cos^2\theta + b\sin^2\theta \quad (2)$$

where $a = 0.64$ and $b = 1.02$ are the fit parameters. Fit parameters in eqs. (2)–(4) reflect the degree of anisotropy of the crystal structure. Detailed calculation of $(l, m, n)$ is described in S3. Compared with the case for along the (102) direction, optical absorption is stronger along the (010) direction for 1PA. This anisotropy can be intuitively understood in terms of optical birefringence between these two crystallographic directions. In fact, our observation is consistent with the previous ellipsometry data,[13] indicating that optical absorption is much more efficient along the (010) direction at our excitation wavelength of $\lambda = 240$ nm. The theory for the 1PA selection rules in the monoclinic structure is detailed in S4 and S5.

In contrast, selection rules for nonlinear optical 2PA ($\lambda = 460$ nm) and 3PA ($\lambda = 550$ nm) are quite different since the axis for maximum absorption is not the (010) direction, but the (102) direction. The 2PA polarization dependence (pink dots) is plotted in Figure 3c, together with the fit in accordance with the theoretical 2PA selection rules (red) given by

$$|a\cos^2\theta + b\sin^2\theta|^2 + |c\cos\theta\sin\theta|^2 \quad (3)$$

where $a = 1$, $b = 3.4$, and $c = 0.4$ are the fit parameters. Compared with the 1PA case, the observed polarization dependence is less intuitive but best explained by the $C_{2h}$ symmetry for 2PA transition, which corresponds to eq. (3). The case for 3PA (blue dots) is shown in Figure 3d, where the low symmetry of the crystal is revealed in the substructure of the observed pattern. The red trace in Figure 3d corresponds to the theoretical fit to the observed 3PA polarization, which is given by

$$|a\cos^3\theta|^2 + |\sin\theta(b\cos^2\theta + c\sin^2\theta) + d\sin\theta\cos^2\theta|^2 \quad (4)$$

where $a = 1$, $b = 13$, $c = 84$ and $d = 13$ are the fit parameters. The theoretical calculations for the 2PA and 3PA selection rules are much more complicated as they are nonlinear optical processes and details are described in S4 and S5.

We further determined the 2PA coefficient ($\beta$) and the 3PA coefficient ($\gamma$) by monitoring the nonlinear absorbance of the input beam by the β-Ga$_2$O$_3$ crystal as a function of input intensity, when $\hat{\epsilon} = (l, m, n)$ was set to parallel to the (102) direction in order to ensure maximum nonlinear absorption (Figure 3c,d). In our experiment, we varied the input intensity $I(\phi)$ by adjusting the angle ($\phi$) of the step-variable ND filter while the input polarization of the excitation laser remained the same. The pink and blue dots in Figure 4 correspond to the input intensity $I_b(\phi)$ vs. output intensity $I_a(\phi)$ for (a) 2PA at λ = 460 nm and (b) 3PA at λ = 550 nm, respectively. With increasing input, the data points start to deviate from the black dashed line that represents the case for the absence of nonlinear absorption ($\beta = 0$ and $\gamma = 0$). The red curve in each plot is the numerical fit to determine the values for $\beta$ and $\gamma$ from the ratio $I_a(\phi)/I_b(\phi)$, corresponding to the normalized transmittance:

$$\frac{I_a(\phi)}{I_b(\phi)} = \frac{2}{\pi^{1/2}p(\phi)} \int_0^\infty \ln[1 + p(\phi)e^{-x^2}] dx \quad (5)$$

$$\frac{I_a(\phi)}{I_b(\phi)} = \frac{2}{\pi^{1/2}q(\phi)} \int_0^\infty \ln[\sqrt{1 + q^2(\phi)e^{-x^2}} + q(\phi)e^{-x^2}] dx \quad (6)$$

Here, the intensity-dependent parameters of $p(\phi)$ and $q(\phi)$ are defined by $p(\phi) = \beta I(\phi)d$ and $q(\phi) = (2\gamma d)^{1/2} I(\phi)$, where $d \sim 0.7$ mm is the sample thickness. In the calculation of $p(\phi)$ and $q(\phi)$, we did not consider the effective thickness as linear absorption at 460 nm or 550 nm is quite negligible. Therefore, eq. 5 or 6 to the data has only a single fit parameter, which is simply $\beta$ or $\gamma$. The best fits were obtained with $\beta = 3.45$ cm GW$^{-1}$ at $\lambda = 460$ nm and $\gamma = 0.013$ cm$^3$ GW$^{-2}$ at $\lambda = 550$ nm, respectively. Our measured $\beta$ value is somewhat larger than the recently reported value using Z scan.[30] Note that the $\beta$ value corresponds to the normalized transmittance of 85% at $I(\phi) = 2.0$ GW cm$^{-2}$ and the $\gamma$ value to the normalized transmittance of 88% at 16.0 GW cm$^{-2}$.

In general, the nonlinear absorption coefficient of a material depends strongly on its bandgap and band structure as well. We compared the measured 2PA and 3PA coefficients with the theoretical values predicted by the two-band model in order to assess β-Ga$_2$O$_3$ as a potential UV nonlinear optical material. As for the 2PA coefficient, we adopted the isotropic parabolic band theory based on the S-matrix formalism,[20,21,31] yielding a wavelength-dependent $\beta(\lambda)$ given by[32–34]

$$\beta(\lambda) = \beta\left(x = \frac{E}{E_g}\right) = K \frac{\sqrt{E_p}}{n_0^2 E_g^3} \frac{(2x-1)^{3/2}}{(2x)^5} \quad \text{(cm GW}^{-1}) \quad (7)$$

where $x$ is the dimensionless dispersion parameter defined by the ratio of photon energy ($E = hc/\lambda$) and bandgap energy ($E_g$) with $h$ and $c$ being the Planck constant and the speed of light at vacuum, respectively. In eq. 7, $n_0$ is the linear refractive index, which is about 2.0 for β-Ga$_2$O$_3$,[13] and $E_p$ is a constant given by $E_p = 2|p_{vc}|^2/m$, where $p_{vc}$ is the interband momentum matrix element and $m$ is the free-electron mass. The widely accepted numerical value for $E_p$ is 21 eV, which is believed to be nearly material independent for dipole-allowed direct-gap semiconductors. However, this may be not the case for highly asymmetric β-Ga$_2$O$_3$, where $p_{vc}$ has a strong directional dependence. In fact, our polarization dependence clearly indicates that $\beta$ has a drastic directional dependence as demonstrated in Figure 3c. Clearly, eq. 7 is valid only within the 2PA band ($1/2 < x < 1$) and the absolute magnitude of the $\beta$ value is determined by an overall scaling factor $K$, which is about 1940 in units such that $\beta$ is in cm GW$^{-1}$ in the two-band model.[31] When plugging our case of $x = (hc/\lambda E_g) = 0.57$ at λ = 460 nm into eq. 7 and assuming $E_p = 21$ eV, the theoretically predicted $\beta$ value is about 0.66 cm GW$^{-1}$, which is approximately 5 times smaller than our experimentally determined value along the (102) direction. This implies that the actual $p_{vc}$ value for β-Ga$_2$O$_3$ should be larger by the same factor in this particular direction. Therefore, our results show that β-Ga$_2$O$_3$ is an excellent two-photon absorber with a drastic polarization dependence, which could be very useful for polarization-selective 2PA spectroscopy and/or microscopy.

The theoretical expression for the wavelength-dependent $\gamma(\lambda)$ can be also calculated within the two-band model, which is given by;[34]

$$\gamma(\lambda) = \gamma\left(x = \frac{E}{E_g}\right) = K' \frac{\sqrt{E_p}}{n_0^3 E_g^6} \frac{(3x-1)^{\frac{5}{2}}}{(3x)^9} \quad \text{(cm}^3 \text{ GW}^{-2}) \quad (8)$$

where $x$ is the dimensionless dispersion parameter defined within the 3PA band ($1/3 < x < 1/2$) and $K'$ is an overall scaling factor, which is about 4140 in units such that $\gamma$ is in cm$^3$ GW$^{-2}$ (see S6 for details). By taking into account the dispersion parameter of β-Ga$_2$O$_3$ at $x = (hc/\lambda E_g) = 0.47$ for λ = 550 nm and the actual value of $p_{vc}$ determined from the 2PA case in eq. 7, eq. 8 yields a theoretical value of $\gamma = 0.008$ cm$^3$ GW$^{-2}$, which is reasonably consistent with our experimental value within a factor of two. This also implies that β-Ga$_2$O$_3$ is an excellent three-photon absorber with the polarization contrast shown in Figure 3d. Even with a large bandgap value of 4.5 eV, the observed strong nonlinear optical properties of β-Ga$_2$O$_3$ stem from its band structure through $p_{vc}$, which takes into accounts the curvatures of the associated valence and conduction bands via their effective masses.

According to the literature,[35,36] β-Ga$_2$O$_3$ is known to exhibit the below-gap PL even if it is excited by light whose energy is somewhat lower than the bandgap, i.e., subgap excitation. This observation raised the issue of the presence of an indirect bandgap[6–8,17] or a deep acceptor level, in which the latter is indeed related to the self-trapped-hole state lying about 1.1 eV above the valence band.[18] In order to clarify the issue, we investigated absorption power dependence as a function of wavelength across the bandgap over the range from 220 nm to 300 nm as displayed in Figure 5. In our power-dependent PLE spectroscopy, $\hat{\epsilon} = (l, m, n)$ was parallel to the (102) direction. Firstly, the critical exponent of $p = 1$ at the bandgap (~4.74 eV) clearly indicates that transition is basically excitonic, but involving the extrinsic V$_O$ and V$_{Ga}$ levels. The same power exponent persists up to ~275 nm, but suddenly increases to 1.4 near 280 nm and further to 1.6 around 300 nm. This intriguing feature can be explained by the subgap 1PA via a significant Urbach tailing of the direct gap to 275 nm (~4.5 eV) and the coexistence of the subgap 1PA and 2PA thereafter, which explains the non-integer exponent in the range of $1 < p < 2$. This indicates that subgap 1PA is only significant within the Urbach tailing range (4.5 eV–4.74 eV). In other words, if this subgap transition is exclusively due to the presence of the self-trapped state, 1PA should persist all the way down to 1.1 eV below the bandgap (3,66 eV–4.74 eV). Moreover, we experimentally confirmed that the subgap 1PA is as efficient as typical band-to-band transition. This also excludes the possibility of the role of the indirect gap as indirect transition is much more inefficient. Moreover, a theoretically predicted indirect gap is only 30 meV–40 meV below the bandgap. Therefore, our power-dependent PLE results show that the subgap transition arises essentially from the Urbach tailing effect. A slight decrease in the $p$ value above the bandgap may be caused by density-dependent effects such as nonradiative exciton Auger decay[37] and/or saturation of the $V_O$ and $V_{Ga}$ sites as optical absorption rapidly increases above the band edge.

Lastly, we carried out 2PA depth scan to examine the spectral feature of the $V_O$ and $V_{Ga}$ transitions as a function of depth from the excitation surface. One of the advantages of 2PA method is the ability to measure the depth-dependent profile by controlling the focus inside the sample.[38] Recently, the variation of the PL shape was demonstrated using DRCL spectroscopy in support of the extrinsic nature for the below-gap PL. Figure 6a shows the ratio of the $V_{Ga}$ transition to the $V_O$ transition under 2PA as function of sample position Z, where the sample was translated from Z = -100 μm (focus in the middle of the sample) to 200 μm (outside of the Rayleigh range). Three representative PL spectra were plotted in Figure 6b, c, d with the schematics for the corresponding Z positions in the insets. When the beam focus is in the middle of the sample, the 2PA-induced PL from the bulk is more dominant over that from the surface (bulk excitation in Figure 6b). When the beam focus is precisely at the surface of the sample, both bulk and surface contributions coexist (intermediate excitation in Figure 6c). If the sample is outside the Rayleigh range, 2PA occurs predominantly at the surface with a negligible bulk contribution (surface excitation in Figure 6d). We emphasize that results in Figure 6 cannot be explained by the polaron picture that should exhibit an intrinsic depth-independent PL. In fact, our 2PA depth scan is consistent with the recent DRCL results.[12] Our observation shows that the origin of the below-gap PL is the extrinsic $V_{Ga}$ and $V_O$ levels that have different density distributions from the surface to the interior of the sample. We note here that a non-uniform compensation in Sn-doped β-$Ga_2O_3$ was reported recently,[39] in which stronger compensation of the donor (Sn) near the surface was observed, compared to the inside of the sample. The suspected candidate for this non-uniform compensation was also attributed to the stronger contribution of $V_{Ga}$ near the surface. Our result further provides a clear experimental evidence of the existence of non-uniform distribution of $V_{Ga}$ and $V_O$ from the surface, which can be a crucial determination factor for the realization of highly conductive n-type and p-type $Ga_2O_3$.

In summary, highly asymmetric optical properties of β-$Ga_2O_3$ have been investigated by employing both linear and nonlinear PLE spectroscopy at room temperature. Along with conventional 1PA, we showed that this wide-gap semiconductor can be nonlinearly excited by 2PA and 3PA, in which optical excitation relaxes predominantly into the ground state via defect-induced PL at 3.0 eV and 3.5 eV. The absorption power dependence and 2PA depth scan clearly show that these below-gap transitions arise essentially from radiative recombination of excitons bound to $V_{Ga}$ and $V_O$ sites, respectively. The strong non-uniform distribution of $V_{Ga}$ and $V_O$ sites from the sample surface to the inside of the sample was observed, which can explain the recently reported stronger compensation of the donor near the surface. We showed that the $V_{Ga}$ PL and the $V_O$ PL are highly polarized along the (102) direction of β-$Ga_2O_3$ regardless of the excitation order. The absorption polarization dependences were also measured to be highly polarized for 1PA, 2PA, and 3PA, respectively, which are quite distinct for each excitation order. The observed absorption polarization dependences were explained theoretically based on optical selection rules for the monoclinic crystal structure. We used the input-output method to determine the 2PA coefficient $\beta$ = 3.45 cm $GW^{-1}$ at λ = 460 nm and the 3PA coefficient $\gamma$ = 0.013 $cm^3$ $GW^{-2}$ at λ = 550 nm, respectively. It turned out that this wide-gap semiconductor possesses at least five times higher nonlinearity when compared with typical two-band models. The excellent nonlinearity of β-$Ga_2O_3$ stems from its highly anisotropic band structure. Our series of important results imply that this wide-gap semiconductor is not only a highly promising UV nonlinear absorber with high polarization contrast, but also a UV light emitter especially when defect control is ensured for the band-edge emission.

**AUTHOR INFORMATION**

**Corresponding Author**

*E-mail (J.I. Jang): jjcoupling@sogang.ac.kr.

ORCID

Joon I. Jang: 0000-0002-1608-8321

**ACKNOWLEDGMENTS**

This work was supported by the Basic Science Research Program (2017R1D1A1B03035539), and was also supported the Basic Science Research Program through the National Research Foundation of Korea (NRF) funded by the Ministry of Education (2017R1D1A3A03000947). The authors acknowledge Dr. Y. Moon at UJL Co. and Prof. J. S. Ha at Chonnam National University for providing α-$Ga_2O_3$ samples and Prof. K.-B. Chung at Dongguk University and Prof. T. J. Kim at Kyung Hee University for preliminary discussions. J. Cho acknowledges Dr. H. R. Byun for helpful discussions on the experiments.

**REFERENCES**

[1]　F. Bechstedt, J. Furthmüller, *Appl. Phys. Lett.* **2019**, *144*, 122101.

[2]　M. Higashiwaki, A. Kuramata, H. Murakami, Y. Kumagai, *J. Phys. D. Appl. Phys.* **2017**, *50*, 333002.

[3]　M. Higashiwaki, H. Murakami, Y. Kumagai, A. Kuramata, *Jpn. J. Appl. Phys.* **2016**, *55*, 1202A1.

[4]　M. A. Mastro, A. Kuramata, J. Calkins, J. Kim, F. Ren, S. J. Pearton, *ECS J. Solid State Sci. Technol.* **2017**, *6*, 356.

[5]　M. Kim, J. H. Seo, U. Singisetti, Z. Ma, *J. Mater. Chem. C* **2017**, *5*, 8338.

[6]　S. J. Pearton, J. Yang, P. H. Cary Iv, F. Ren, J. Kim, M. J. Tadjer, M. A. Mastro, *Appl. Phys. Rev.* **2018**, *5*, 011301.

[7]　S. I. Stepanov, V. I. Nikolaev, V. E. Bougrov, A. E. Romanov, *Rev. Adv. Mater. Sci.* **2016**, *44*, 63.

[8]　H. Peelaers, J. B. Varley, J. S. Speck, C. G. Van De Walle, *Appl. Phys. Lett.* **2018**, *112*, 242101.

[9]　J. Kim, J. Jang, *Aerosol Sci. Technol.* **2018**, *52*, 557.

[10]　R. Jangir, S. Porwal, P. Tiwari, P. Mondal, S. K. Rai, T. Ganguli, S. M. Oak, S. K. Deb, *J. Appl. Phys.* **2012**, *112*, 034307.

[11]　W. Mi, C. Luan, Z. Li, C. Zhao, X. Feng, J. Ma, *Opt. Mater.* **2013**, *35*, 2624.

[12]　H. Gao, S. Muralidharan, N. Pronin, M. R. Karim, S. M. White, T. Asel, G. Foster, S. Krishnamoorthy, S. Rajan, L. R. Cao, M. Higashiwaki, H. Von Wenckstern, M.

Grundmann, H. Zhao, D. C. Look, L. J. Brillson, *Appl. Phys. Lett.* **2018**, *112*, 242102.

[13] T. Onuma, S. Saito, K. Sasaki, T. Masui, T. Yamaguchi, T. Honda, A. Kuramata, M. Higashiwaki, *Jpn. J. Appl. Phys.* **2016**, *55*, 1202B2.

[14] K. Yamaguchi, *Solid State Commun.* **2004**, *131*, 739.

[15] B. E. Kananen, L. E. Halliburton, K. T. Stevens, G. K. Foundos, N. C. Giles, *Appl. Phys. Lett.* **2017**, *110*, 202104.

[16] Q. D. Ho, T. Frauenheim, P. Deák, *Phys. Rev. B* **2018**, *97*, 115163.

[17] K. A. Mengle, G. Shi, D. Bayerl, E. Kioupakis, *Appl. Phys. Lett.* **2016**, *109*, 212104.

[18] S. Yamaoka, M. Nakayama, *Phys. Status Sol. C* **2016**, *13*, 93.

[19] S. Yamaoka, Y. Mikuni, M. Nakayama, *J. Phys. Soc. Jpn.* **2019**, *88*, 113701.

[20] A. Pasquarello, A. Quattropani, *Phys. Rev. B* **1991**, *43*, 3837.

[21] M. Inoue, Y. Toyozawa, *J. Phys. Soc. Jpn* **1965**, *20*, 363.

[22] A. R. Hassan, R. Raouf, *Phys. Status Sol.* **1981**, *104*, 703.

[23] T. R. Bader, A. Gold, *Phys. Rev.* **1968**, *171*, 997.

[24] F. O. Saouma, D. Y. Park, S. H. Kim, M. S. Jeong, J. I. Jang, *Chem. Mater.* **2017**, *29*, 6876.

[25] J. B. Varley, H. Peelaers, A. Janotti, C. G. Van De Walle, *J. Phys. Condens. Matter* **2011**, *23*, 334212.

[26] J. B. Varley, J. R. Weber, A. Janotti, C. G. Van De Walle, *Appl. Phys. Lett.* **2010**, *97*, 142106.

[27] N. E. Hsu, W. K. Hung, Y. F. Chen, *J. Appl. Phys.* **2004**, *96*, 4671.

[28] K. Hönes, M. Eickenberg, S. Siebentritt, C. Persson, *Appl. Phys. Lett.* **2008**, *93*, 092102.

[29] J. Liu, Y. Zhao, Y. J. Jiang, C. M. Lee, Y. L. Liu, G. G. Siu, *Appl. Phys. Lett.* **2010**, *97*, 231907.

[30] H. Chen, H. Fu, X. Huang, J. A. Montes, T.-H. Yang, I. Baranowski, Y. Zhao, *Opt. Express* **2018**, *26*, 3938.

[31] M. Sheik-Bahae, D. C. Hutchings, D. J. Hagan, E. W. Van Stryland, *IEEE J. Quantum Electron.* **1991**, *27*, 1296.

[32] M. Sheik-Bahae, D. J. Hagan, E. W. Van Stryland, *Phys. Rev. Lett.* **1990**, *65*, 96.

[33] A. A. Said, M. Sheik-Bahae, D. J. Hagan, T. H. Wei, J. Wang, J. Young, E. W. Van Stryland, *J. Opt. Soc. Am. B* **1992**, *9*, 405.

[34] H. S. Brandi, C. B. De Araujos, *J. Phys. C Solid State Phys.* **1983**, *16*, 5929.

[35] G. Pozina, M. Forsberg, M. A. Kaliteevski, C. Hemmingsson, *Sci. Rep.* **2017**, *7*, 42132.

[36] Y. Berencén, Y. Xie, M. Wang, S. Prucnal, L. Rebohle, S. Zhou, *Semicond. Sci. Technol.* **2019**, *34*, 035001.

[37] J. I. Jang, J. P. Wolfe, *Phys. Rev. B* **2005**, *72*, 241201.

[38] T. Tanikawa, K. Ohnishi, M. Kanoh, T. Mukai, T. Matsuoka, *Appl. Phys. Express* **2018**, *11*, 031004.

[39] A. Y. Polyakov, N. B. Smirnov, I. V. Shchemerov, D. Gogova, S. A. Tarelkin, S. J. Pearton, *J. Appl. Phys.* **2018**, *123*, 115702.

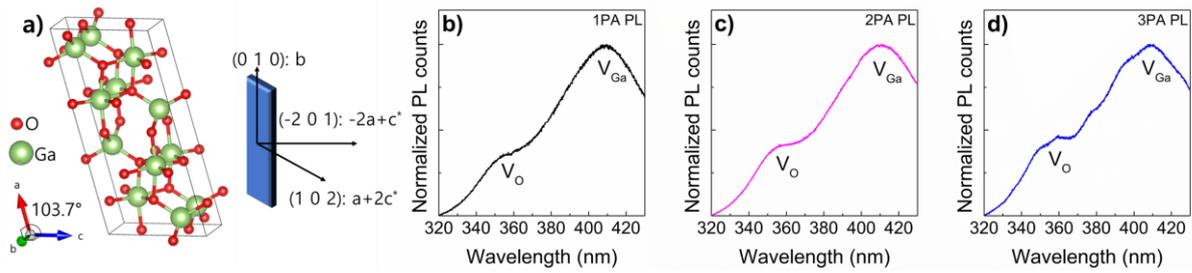

**Figure 1.** (a) Crystal structure of β-Ga$_2$O$_3$ and excitation geometry for optical measurements. Normalized PL spectra from β-Ga$_2$O$_3$ at room temperature when excited at (b) λ = 240 nm (1PA), (c) λ = 290 (2PA), and (d) λ = 550 nm (3PA), respectively. The peak assignment was made in accordance with Ref. [12].

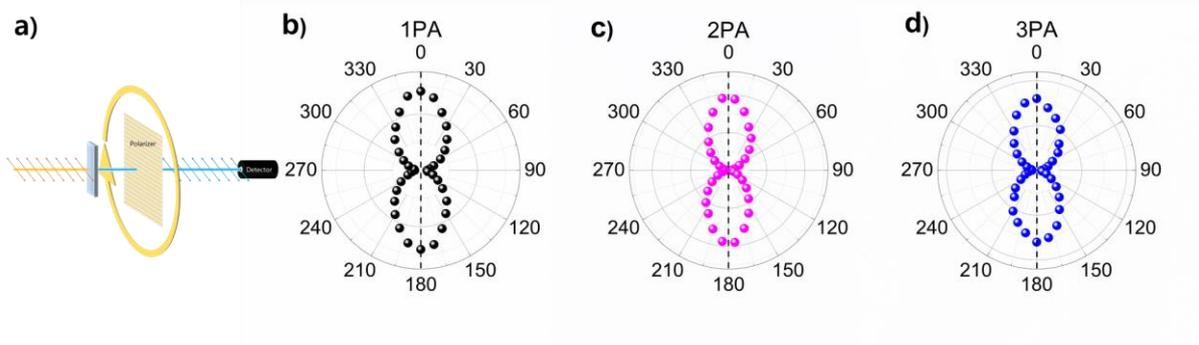

**Figure 2.** (a) Schematic for measuring the PL polarization when the input polarization is parallel to the (102) direction. Polarization dependence of the $V_{Ga}$ transition plotted in the polar coordinate system for (b) 1PA ($\lambda$ = 240 nm), (c) 2PA ($\lambda$ = 460 nm), and (d) 3PA ($\lambda$ = 550 nm), respectively. The (102) direction is along the line connecting 0° and 180°.

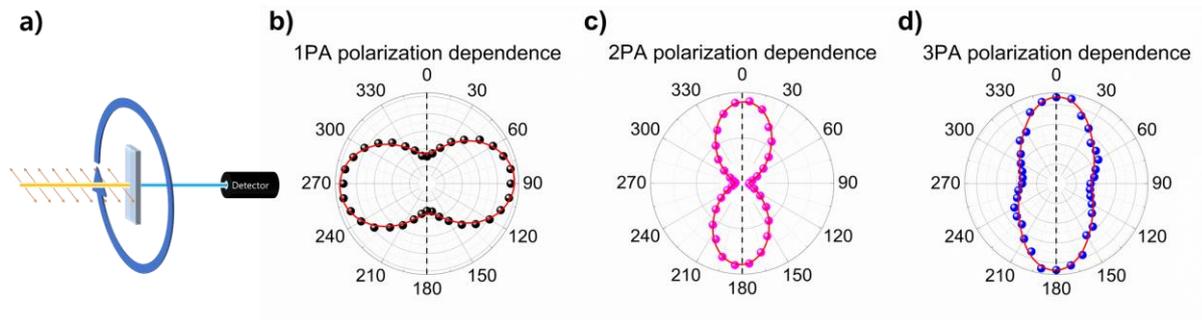

**Figure 3.** (a) Schematic for measuring the absorption polarization dependence. Optical selection rules plotted in the polar coordinate system for (b) 1PA at λ = 240 nm, (c) 2PA at λ = 460 nm, and (c) 3PA at λ = 550 nm, respectively. The (102) direction is along the dashed line connecting 0° and 180°. The superimposed theoretical fits (red) were obtained from eqs. (2)–(4).

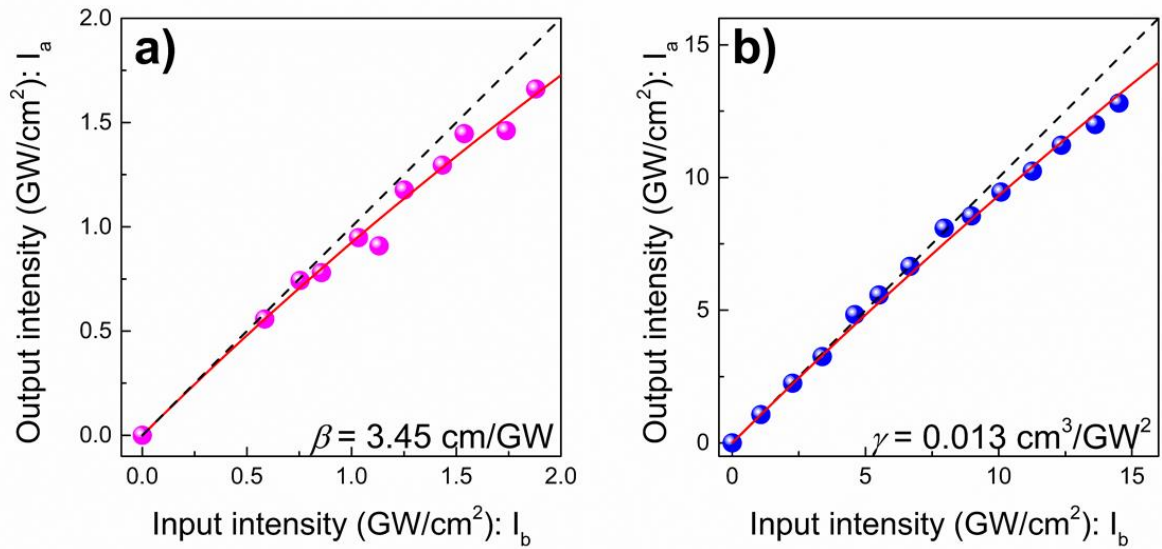

**Figure 4.** $I_b(\phi)$ vs. $I_a(\phi)$ at (a) λ = 460 nm (2PA) and (b) λ = 550 nm (3PA). The red curves indicate the best fits using eqs. 5 and 6, yielding $\beta$ = 3.45 cm GW$^{-1}$ and $\gamma$ = 0.013 cm$^3$ GW$^{-2}$, respectively. The black dashed lines correspond to the case for $I_a(\phi) = I_b(\phi)$ (no nonlinear absorption).

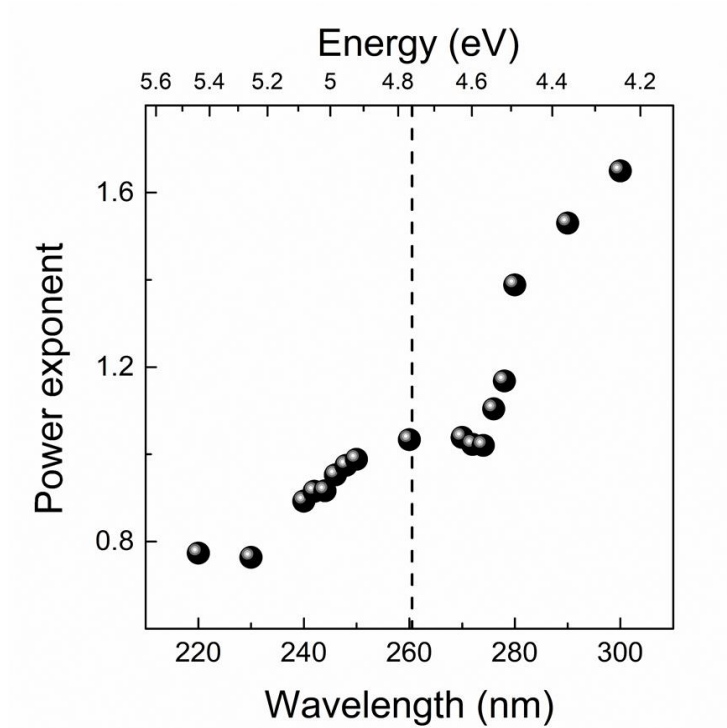

**Figure 5.** Excitation wavelength vs. critical exponent of the absorption power dependence across the bandgap (4.74 eV). The critical exponent near the band edge is about 1.0, basically consistent with our model of the bound excitonic PL. A slight decrease in the exponent in the high-energy end seems to arise from density-dependent effects, whereas both 1PA and 2PA contribute to optical excitation for λ in the range from 270 nm–300 nm.

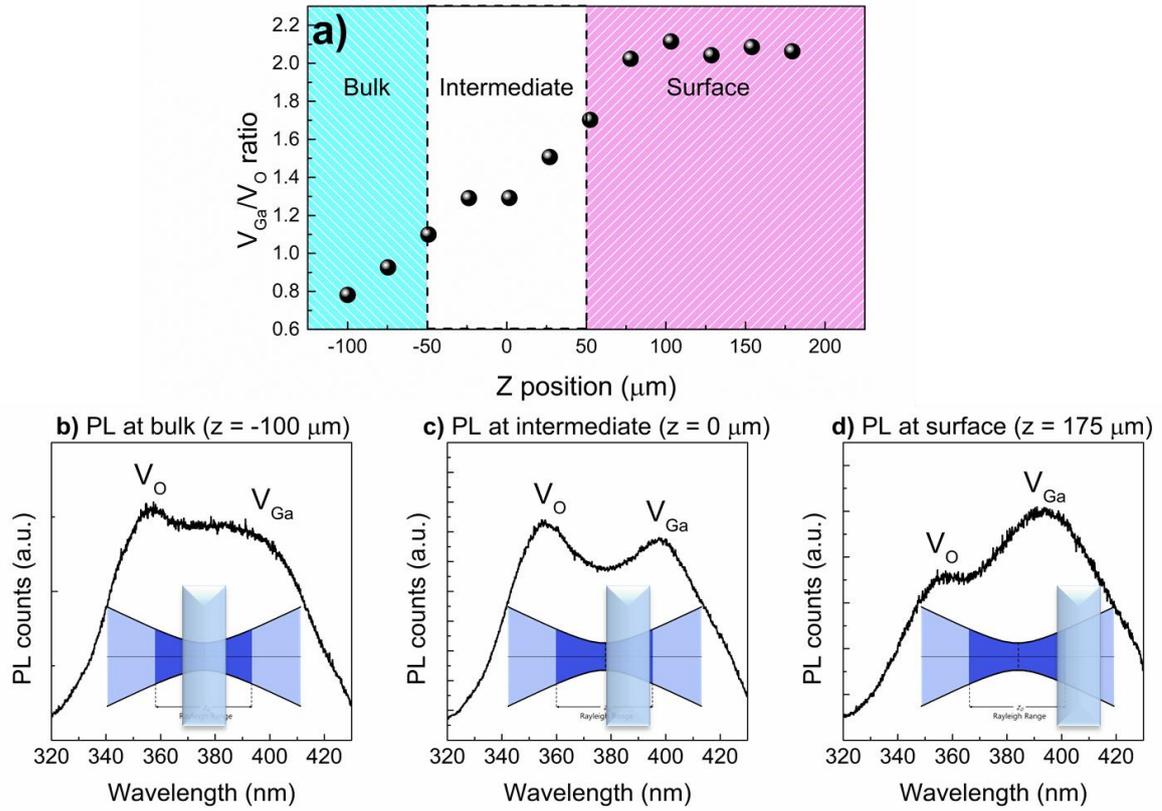

**Figure 6.** (a) Z position vs. $V_{Ga}/V_O$ ratio, showing the depth variation of each vacancy density. The PL spectra at (b) Z = –100 μm (bulk excitation), (c) Z = 0 μm (intermediate excitation) (d) and Z = 175 μm (surface excitation), respectively. The inset in each spectrum schematially illustrates the excitation condition as a function of Z. A clear change in the PL spectral shape strongly supports that the origin of the PL is extrinsic $V_{Ga}$ and $V_O$ whose densities vary differently as a function of depth.